\documentclass[prl,twocolumn,showpacs,superscriptaddress]{revtex4}
\usepackage{graphicx}
\begin{document}
\title{Phase dynamics after connection of two separate
Bose-Einstein condensates
}

\author{I. Zapata}
\altaffiliation[Present address:]{ Neworldcity, C/ Almagro 2,
    E-28010 Madrid, Spain}
\affiliation{Department of Physics, University of Illinois at
    Urbana-Champaign, 1110 West Green Street, Urbana, IL 61801}
\affiliation{Departamento de F\'{\i}sica Te\'orica de la Materia Condensada
    and Instituto de Ciencia de Materiales ``Nicol\'as Cabrera'',
    Universidad Aut\'onoma de Madrid, E-28049 Madrid, Spain}
\author{F. Sols}
\affiliation{Departamento de F\'{\i}sica Te\'orica de la Materia Condensada
    and Instituto de Ciencia de Materiales ``Nicol\'as Cabrera'',
    Universidad Aut\'onoma de Madrid, E-28049 Madrid, Spain}
\author{A. J. Leggett}
\affiliation{Department of Physics, University of Illinois at
    Urbana-Champaign, 1110 West Green Street, Urbana, IL 61801}
\date{\today}
%
\begin{abstract}
We study the dynamics of the relative phase following the
connection of two independently formed Bose-Einstein condensates.
Dissipation is assumed to be due to the creation of quasiparticles
induced by a fluctuating condensate particle number. The coherence
between different values of the phase, which is characteristic of
the initial Fock state, is quickly lost after the net exchange of
a few atoms has taken place. This process effectively measures the
phase and marks the onset of a semiclassical regime in which the
system undergoes Bloch oscillations around the initial particle
number. These fast oscillations excite quasiparticles within each
condensate and the system relaxes at a longer time scale until it
displays low-energy, damped Josephson plasma oscillations,
eventually coming to a halt when the equilibrium configuration is
finally reached.

\end{abstract}
\maketitle


%

The way in which two Bose condensates, originally separately
formed, behave when they are connected poses a fundamental problem
in non-equilibrium macroscopic quantum mechanics. Two separate
condensates 1 and 2 can be described by a relative number
eigenstate $|n \rangle$ which in the phase representation may be
written \cite{comment1}
\begin{equation}
|n\rangle = \int_{0}^{2\pi} \frac{d\varphi}{\sqrt{2\pi}}
e^{in\varphi} |\varphi \rangle \label{Fockstate},
\end{equation}
where $|\varphi \rangle$ is an eigenstate of the relative phase
\cite{comment11,legg01} $\varphi=\varphi_1-\varphi_2$, with
normalization
$\langle\varphi|\varphi'\rangle=\delta(\varphi-\varphi')$, and
$n=(N_1-N_2)/2$ is the number of transferred atoms. The Fock state
(\ref{Fockstate}) may be viewed as resulting from a coherent
superposition of all possible values of the relative phase. This
poses no problem as long as the condensates are physically
separated, e.g. by a barrier so high as to be effectively
impenetrable, since then no physical observable depends on
$\varphi$. The situation changes when the two condensates are
brought into connection, e.g. by lowering the barrier to the point
where Josephson tunnelling becomes appreciable. Then a nonzero
supercurrent develops which depends on the phase through the
relation
\begin{equation}\label{current-phase}
\dot{n}=I(\varphi)=(E_J/\hbar)\sin\varphi,
\end{equation}
where $E_J$ is the Josephson coupling energy. If we interpret the
Josephson relation (\ref{current-phase}) as an operator identity,
we conclude that a system initially prepared in the state
(\ref{Fockstate}) will display a coherent superposition of states
with macroscopically different values of the Josephson current
(\ref{current-phase}). After some time, these different values of
the current will give rise to macroscopically different values of
the relative particle number
\begin{equation}\label{particlenumber}
n(t)=n_0+\int_0^t I \, dt.
\end{equation}
The resulting macroscopic coherence is fragile and likely to be
quickly lost by a tiny perturbation from the environment. After
decoherence sets in, the relative phase is effectively measured
and the system is expected to be describable by a reduced density
matrix that is asymptotically diagonal in the phase
representation. We will show that such a density matrix is a
uniform mixture of pure state density matrices, each of them lying
in the semiclassical regime where phase and number may be taken as
simultaneously well-defined. In general the system finds itself
very far from equilibrium and, due to the interaction with the
dissipative environment, it eventually relaxes to the equilibrium
configuration.

The purpose of this work is to analyze such a dynamical process.
We assume that dissipation is caused by the spontaneous creation
of quasiparticles within each well due to a fluctuating number of
particles in the condensate that is caused in turn by the coherent
exchange of atoms. Such an exchange can, of course, occur only in
the presence of a physical coupling, e.g. of the Josephson type,
between the two condensates. When the Josephson coupling is
operating, the Hamiltonian of the system at low energies may be
written \cite{zapa98,sols99}
\begin{equation}
H = E_J(1-\cos\varphi)+E_c n^2 /2,
\label{JosephsonHamiltonian}
\end{equation}
where $E_c$ represents (up to a constant) the effects of the
interatomic interaction for small values of $n$. When $E_J$
becomes nonzero, $n$ is no longer a constant of motion, as
explicitly indicated by the equation of motion
(\ref{current-phase}), and its dynamic fluctuations will induce
excitation of quasiparticles within each well. The description of
such a process will require going beyond the pendulum Hamiltonian
(\ref{JosephsonHamiltonian}) by including the dissipative coupling
to the quasiparticle field. In practice, there may be other
dissipative mechanisms, such as incoherent exchange of single
atoms \cite{zapa98} or coupling to the walls of a cavity
\cite{jaks01}, but quasiparticle creation is an intrinsic
mechanism which lends itself to an approximate analytical
treatment.

The problem which we address in this article is fundamental in
several respects. First, it presents an important case of (phase)
quantum measurement that can be treated in detail. Second, the
connection between independent BEC's has the essential features of
a macroscopic interference experiment
\cite{andr97,ande98,ande84,legg91,cast97,kohl01}, since it
involves an initially random relative phase which becomes well
defined, in this case due to the introduction of a Josephson
coupling with the help of the quasiparticle field. Finally, the
establishment of phase coherence between two recently connected
condensates has bearing on the ``Kibble problem'' of formation of
cosmological singularities in the early stages of the Universe
\cite{kibb76}. Recent work on condensate coherence in optical
lattices \cite{grei02} suggests that the study of phase dynamics
after nonadiabatic switching of the Josephson coupling is within
experimental reach.

The analytical treatment which we present in this article requires
the introduction of some approximations. We have already said that
dissipation is assumed to be caused by the creation of intrawell
quasiparticles caused by a fluctuating condensate number.
Following ref. \cite{meie00}, we take box normalization for the
one-atom wave functions within each well. This is not an important
limitation in the context of laser confined alkali condensates.
The initial relative particle number is assumed to be sufficiently
large for the semiclassical approximation to be applicable just
after decoherence takes over between distant phase values. We also
assume that the two independently prepared condensates are
identical and neglect quasiparticle tunneling. For conceptual
simplicity, we will focus here on the case in which macroscopic
quantum self-trapping takes place \cite{smer97,milb97}.  This
requires $E_c n_0^2 > 4 E_J$. In the future it will be desirable
to extend the present analysis by relaxing some of these
approximations.

\begin{figure}[b]
\includegraphics[height=.95\columnwidth,angle=-90]{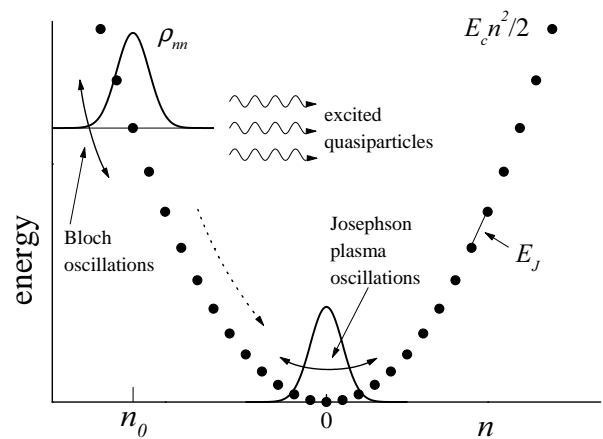}
\caption{\label{schematic} Schematics of the connection process in
the number representation: After a time $t\sim 1/\gamma_d$ a
semiclassical wave packet forms that undergoes Bloch oscillations
thus exciting quasiparticles. For $t \agt 1/\gamma_r$, the system
relaxes its energy, eventually performing underdamped Josephson
plasma oscillations. }
\end{figure}

The picture that emerges from our theoretical study is
schematically depicted in (Fig.~\ref{schematic}). There are at
least three time scales in the connection process. First,
decoherence between different phase values takes place after a few
atoms have been exchanged. This process effectively measures the
macroscopic phase. The second time scale is the Bloch oscillation
period around the initial value of the relative particle number.
These oscillations are a manifestation of the ac Josephson effect
due to a chemical potential difference. Finally, the system
relaxes in a longer time scale towards the equilibrium
configuration. Damping is due to the irreversible energy transfer
from the quickly oscillating number-phase degree of freedom to the
quasiparticle field within each well.

First we perform a semiclassical study in which the relative phase
and number are taken as $c$-numbers while the quasiparticle field
is treated quantum mechanically. The semiclassical assumption is
later justified from a full quantum treatment which explicitly
shows how the quasiparticle field transforms the reduced density
matrix of the phase-number coordinate from that of a pure Fock
state to a mixture of Gaussian wave packets. This is the process
that effectively ``measures'' the phase and makes it well-defined
in a particular experimental realization.

At the semiclassical level (where the question of identity or not
between the states $|\varphi\rangle$ and $|\varphi+2\pi\rangle$
can be neglected), it is clear from Eq.
(\ref{JosephsonHamiltonian}), that the problem maps on to that of
the the ``Bloch oscillations'' of a particle in a tight-binding
lattice which is subject to a harmonic potential, with $n$ playing
the role of (discrete) coordinate and $\varphi$ that a of a wave
vector. A simple semiclassical analysis reveals that a particle
placed with initial ``wave vector'' $\varphi_0$ at a high point
$n_0$ of the harmonic tight-binding lattice, undergoes periodic
motion.
Specifically, for $E_c n_0^2 \gg E_J$, and with an appropriate
choice of time origin, we may write
\begin{equation} \label{n-dynamic}
n(t)=\bar{n}+p \sin(\omega_0 t) ,
\end{equation}
where $p\equiv(E_J/\bar{n} E_c)$, $\omega_0\equiv
E_c\bar{n}/\hbar$, and, initially, $\bar{n} \simeq n_0$. Equation
(\ref{n-dynamic}) describes Bloch oscillations undergone by a
quantum particle in a tilted lattice. The resulting self-trapping
is not stable because the rapidly oscillating $n(t)$ induces
excitations in the quasiparticle field, which acts as a dissipate
environment for the macroscopic degree of freedom
\cite{ruos98,legg77}. These oscillations are very fast and one is
rather interested in the slower dynamics of the time average
$\bar{n}$ as it experiences damping.

Since quasiparticles within each well are sensitive to the number
of particles in the condensate to the extent that they are but
linear deviations from the mean field description \cite{dalf99},
it is clear that (\ref{n-dynamic}) yields a time-dependent
perturbation that will excite quasiparticles. The time-dependent
Hamiltonian in e.g. the left well is \cite{fett71}
\begin{eqnarray}\label{t-Hamiltonian}
H(t)=\sum_k \left[\varepsilon_k+\frac{g}{V}\left(\frac{N}{2}+n(t)
\right)\right]c_k^{\dagger}c_k \\
+ \sum_k \frac{g}{2V}\left(\frac{N}{2}+n(t)
\right)(c_{k}^{\dagger} c_{-k}^{\dagger}+ c_{-k} c_{k}),
\end{eqnarray}
where $g$ is the coupling constant, $N$ is the total particle
number, and $V$ is the volume of one box. The Hamiltonian for the
right well looks like (\ref{t-Hamiltonian}) with $n(t)$ replaced
by by $-n(t)$. Eq. (\ref{t-Hamiltonian}) may be rewritten as
\begin{equation}\label{dynamic-perturbation}
H(t)=H_0+A p\sin(\omega t),
\end{equation}
where $H_0$ is the unperturbed ``bath'' Hamiltonian [with $n(t)$
replaced by $\bar{n}$] and
\begin{equation}\label{bath-operator}
A=\frac{g}{2V}\sum_k (2c_{k}^{\dagger}c_{k}+ c_{k}^{\dagger}
c_{-k}^{\dagger} + c_{-k} c_{k}).
\end{equation}
By applying linear response theory \cite{forster} with $\bar{n}$
treated as constant, we obtain for the dissipated power within one
well:
\begin{equation}\label{dissipation}
W=(p^2/2)\omega_0 \chi ''(\omega_0),
\end{equation}
where
\begin{eqnarray}\label{chi}
\chi ''(\omega) &\equiv& \frac{1}{2\hbar}\int dt \, e^{i\omega t}
\langle [A(t),A(0)] \rangle \\
\nonumber &=& \frac{\pi g^2}{2\hbar V^2} \sum_k |u_k-v_k|^4 \coth
\left(\frac{\hbar\omega_k}{2k_BT}\right) \delta(\omega-2\omega_k)
\\ \nonumber
&\simeq& \frac{\hbar V}{128 \pi N^2 c^3} \omega^4 \coth
\left(\frac{\hbar\omega}{4k_B T}\right).
\end{eqnarray}
In the third line, we have taken the low frequency approximation
$\hbar\omega \ll mc^2$, $c$ being the sound velocity and $m$ the
atom mass \cite{fett71}. Here $u_k$ and $v_k$ are the coherence
factors of the Bogoliubov transformation, so that, in this limit
\cite{fett71}, $u_k-v_k=(\hbar V ck/Ng)^{1/2}$, where $N/2V$ is
the atom density in one well. We note that $\chi''(\omega)$ goes
like $\omega^4$ at low temperatures and like $\omega^3$ at high
temperatures. The second line of Eq. (\ref{chi}) indicates that
the quasiparticles are excited in pairs, which can be ascribed to
the pairing structure of the depletion cloud \cite{fett72,legg01}.

The energy that is dissipated into the quasiparticle field has to
be taken from the energy of the macroscopic pendulum
(\ref{JosephsonHamiltonian}) which, under the self-trapping
assumption, is dominated by the interaction term. Thus, we write
\begin{equation}\label{energy-balance}
2W= - E_c \bar{n} \dot{\bar{n}},
\end{equation}
where the factor of 2 accounts for the presence of two wells. We
derive the differential equation
\begin{equation}\label{n-bar-dot}
\dot{\bar{n}}=-C\bar{n}^2\coth\left(\frac{E_c\bar{n}}{4k_B
T}\right),
\end{equation}
where $C\equiv 2\pi\eta E_J^2/\hbar N^4 E_c$, with $E_c=2g/V\equiv
8\pi\hbar^2a/mV$ and $\eta \equiv (Na^3/2\pi V)^{1/2}$ the small
parameter of the dilute gas. At low temperatures ($k_BT\ll
E_c\bar{n}$) we get
\begin{equation}\label{n-low-temp}
\bar{n}(t)=\frac{n_0}{1+\gamma_r t},
\end{equation}
while, at high temperatures ($k_BT\gg E_c\bar{n}$),
\begin{equation}\label{n-high-temp}
\bar{n}(t)=n_0 \exp (-\gamma_r t).
\end{equation}
In (\ref{n-low-temp}) and (\ref{n-high-temp}), $\gamma_r$ is,
respectively, the low and high temperature limit of the general
expression
\begin{equation}\label{gamma-R}
\gamma_r=\frac{\hbar C}{E_c}\omega_0 \coth
\left(\frac{\hbar\omega_0}{4k_B T}\right),
\end{equation}
where $\omega_0$ refers to $n_0$.

As the dissipation process continues, the energy of the
macroscopic pendulum decreases until it eventually undergoes low
energy harmonic oscillations. In this limit, the analysis of the
dissipation process is similar to that described above, with
$\omega_0$ replaced by the Josephson plasma frequency $\omega_{\rm
JP}=\sqrt{E_JE_c}/\hbar$ and with the pendulum energy written as
$\tilde{n}^2E_c/2$, where $\tilde{n}$ is the amplitude of the
number oscillations. Since $\omega_{\rm JP}$ does not depend on
$\tilde{n}$ for $\tilde{n} \ll N$, we obtain for both low and high
temperatures
\begin{equation}\label{n-tilde-t}
\tilde{n}(t)=\tilde{n}(0) \exp (-\gamma_r t),
\end{equation}
$\gamma_r$ being given by (\ref{gamma-R}) with $\omega_0$ replaced
by $\omega_{\rm JP}$.

It remains to justify the semiclassical approximation. For that we
have performed a full quantum calculation of the phase dynamics
right after the connection has been established. The direct
reversal process $n \rightarrow -n$ may be safely neglected. We
generalize quantum dissipation techniques \cite{weis99} to the
case of quadratic coupling in the bath coordinates [see Eq.
(\ref{bath-operator})]. After some lengthy algebra, we find that,
in the phase representation, the reduced density matrix simply
evolves as
\begin{equation}\label{reduced}
\left|\rho(\varphi,\varphi';t)\right|=\frac{1}{2\pi}\exp\left[-4\gamma_d
t \sin^2\left(\frac{\varphi-\varphi'}{2}\right)\right],
\end{equation}
where
\begin{equation}\label{gamma-D}
\gamma_d=\gamma_r n_0 \coth\left( \frac{E_c n_0}{2 k_B T},
\right).
\end{equation}
For $\gamma_d t \gg 1$, Eq. (\ref{reduced}) becomes effectively
equivalent to
\begin{equation}\label{asymptotic}
|\rho(\varphi,\varphi';t)|=(2\pi)^{-1}\exp[- (\varphi-\varphi')^2
\gamma_d t],
\end{equation}
and hence $\gamma_d$ is identified as the decoherence rate. It is
possible to prove that the reduced density matrix
(\ref{asymptotic}) is a mixture of pure state Gaussians $\sim
\exp[-2(\varphi-\theta)^2\gamma_d t]$ with $\theta$ uniformly
distributed between 0 and $2\pi$. Thus we assert that the density
matrix (\ref{asymptotic}) describes a quantum system in which the
variable $\varphi$ is measured with precision $\sim (\gamma_d
t)^{-1/2}$ with the average result chosen among a menu of
uniformly distributed possible values. Here the role of the
``measurement apparatus'' is played by the quasiparticle field
with its many degrees of freedom.

At low temperatures, we obtain $\gamma_d/\gamma_r=n_0$. Now we
note that $2\gamma_r$ is the relaxation of the total energy $E_c
n_0^2/2$ and that $E_c n_0$ is the variation in energy due to the
exchange of a single atom. We conclude that the ratio between the
decoherence and the energy relaxation rates is the ratio between
the total energy and the energy variation associated to the
transfer of one atom. Thus decoherence is achieved as soon as
$|\bar{n}(t)|$ decreases by a few units \cite{comment12}. Once
this has happened, Eq. (\ref{asymptotic}) applies and the system
may be treated semiclassically with an initial value of the phase
that is chosen at random and which is defined with increasing
precision as time evolves \cite{kohl01,comment3}. In other words,
the decoherence rate can be identified with the rate of net atom
exchange. On the contrary, an appreciable energy relaxation
requires the exchange of many atoms. This picture is consistent
with the idea that the net exchange of a single atom plays the
role of the excitation of one quantum of the dissipative bath
\cite{cald85}.

Once we can describe our system as a mixture of semiclassical wave
packets, linear evolution applies to these even in the presence of
dissipation. Since the translational motion of a Gaussian wave
packet is semiclassical, we apply the method initially described
with the confidence that it has been justified by a full quantum
treatment.

The final picture of the connection process is one which involves
several time scales. Decoherence in the phase representation sets
in after the exchange of a few atoms has been completed. For a
given experimental run this can be interpreted as the measurement
of the phase in which a given value is chosen with an accuracy
that improves with time. After the phase is defined, a
semiclassical treatment suffices to conclude that the macroscopic
degree of freedom will undergo very fast Bloch oscillations that
will dissipate energy into the many degrees of freedom of the
quasiparticle field. Ultimately this Bloch oscillations become low
energy Josephson plasma oscillations that are also damped by the
quasiparticle field. At long times we expect that the system
density matrix will be the thermal equilibrium one which
corresponds to a total energy of $E_c n_0^2 / 2$. For large $V$
the corresponding temperature is much smaller than the critical
temperature.

The main message is that the depletion cloud, described by the
quasiparticle field, can be an intrinsic agent for the decoherence
and measurement of the macroscopic phase. We have studied the most
fundamental intrinsic mechanism. Incoherent exchange of thermal
atoms \cite{zapa98} and coupling to the walls of a cavity
\cite{jaks01} will yield additional contributions to both the
decoherence and the energy relaxation rates.

%
This work has been supported by NSF Grant no. DMR-99-86199, by the
Direcci\'on General de Investigaci\'on Cient\'{\i}fica y T\'ecnica
under Grants No.\ PB96-0080-C02 and BFM2001-0172, and by the
Ram\'on Areces Foundation. One of us (I.Z.) acknowledges support
from the Fulbright Commission and Spain's Ministerio de
Educaci\'on y Ciencia.

\begin{thebibliography}{19}
\bibitem{comment1}
One may allow for a classical uncertainty in the knowledge of the
initial relative particle number by describing the system by an
incoherent mixture of Fock states, $\rho=\sum_n |n \rangle
\rho_{nn} \langle n|$.
\bibitem{comment11}
We will neglect throughout this paper effects which tend to zero
in the limit $N \rightarrow \infty$, cf. Ref. \cite{legg01}, sec.
VII.C.
\bibitem{comment12} This suggests that, in principle, a fine measurement of $n$
at different times in various runs would reveal $\gamma_d$ if
$n_0$ is known precisely.
\bibitem{comment3} Eventually the phase spread $\sim
1/\sqrt{\gamma_d t}$ will saturate to a finite value due to
interactions, which cause quantum fluctuations of the phase that
tend to oppose definition in a given experimental run
\cite{kohl01}. We neglect this effect here.
\end{thebibliography}
\end{document}